\documentstyle[aps,epsf]{revtex}  

\def\casb
{\mbox{$\Xi^0 \rightarrow \Sigma^+\,e^-\,\overline{\nu}_e\,$}}

\def\casmub
{\mbox{$\Xi^0 \rightarrow \Sigma^+ \mu^- \bar{\nu}_{\mu}$}}

\def\sigp{\mbox{$\Sigma^+ \rightarrow  p\,\pi^0\,$}}

\def\casn
{\mbox{$\Xi^0 \rightarrow \Lambda\,\pi^0\,$}}

\def\kppp
{\mbox{$K^0_L\rightarrow  \pi^+\,\pi^-\,\pi^0\,$}}

\def\ket
{\mbox{$K^{0}_L\rightarrow \pi^\pm\,e^\mp\,\nu_e$\,}}

\def\ketr
{\mbox{$K^{0}_L\rightarrow \pi^\pm\,e^\mp\,\nu_e \gamma\,$}}

\def\kef
{\mbox{$K^{0}_L\rightarrow \pi^0\,\pi^\pm\,e^\mp\,\nu_e\,$}}

\def\lamb
{\mbox{$\Lambda \rightarrow p\,e^-\,\overline{\nu}_e\,$}}

\def\lamn
{\mbox{$\Lambda \rightarrow p\,\pi^-\,$}}

\def\pizgg
{\mbox{$\pi^{0}\rightarrow \gamma \gamma\,$}}

\begin{document}        

\baselineskip 14pt
\title{Branching Ratio and Form Factor Measurements of $\Xi^0$ Beta Decay}
\author{Ashkan Alavi-Harati (for the KTeV collaboration)}
\address{University of Wisconsin, Madison, Wisconsin 53706}
%
\maketitle              

\begin{abstract}        

We present a branching ratio measurement for the beta decay
of neutral Cascade hyperon, \casb, using the KTeV detector at Fermilab. We 
used the principal decay mode of \casn where \lamn, 
as the flux normalization mode. 
The status of the measurement of the ratio of the 
axial-vector to vector coupling ($g_1$/$f_1$) for the Cascade beta decay 
will also be discussed.
Furthermore, we present the preliminary  branching ratio measurement 
for the muonic channel \casmub.

\end{abstract}   	

\section{Introduction}               


We report a measurement of $\Xi^0$ beta decay, \casb, 
branching ratio  (BR) based on data collected during E799-II
data-taking in Summer of 1997. The first observation and BR measurement
of this decay mode was reported by the KTeV collaboration earlier
~\cite{paper}. 
Under  $d$ and~$s$ quark  interchange, this process is
the  direct  analogue of  the neutron beta
decay, $n  \rightarrow p\,e^-\,\overline{\nu}_e$.  Thus, in the flavor
symmetric quark model, 
differences between   these  two decays arise only from the
differing particle masses  and  from the relevant Cabibbo-Kobayashi- 
Maskawa (CKM)~\cite{cabibbo}
matrix elements ($V_{us}$     rather     than   $V_{ud}$). In
the symmetry  limit, the predicted branching ratio is $(2.61
\pm 0.11)  \times 10^{-4}$. Flavor symmetry violation 
effects~\cite{fsv},\cite{garcia} are expected to modify this branching 
ratio by as much as 20-30\%. The directly-measurable final state 
$\Sigma^+$ polarization will allow measurements  
of form factors, providing additional information on flavor symmetry.

\section{Neutral Hyperon Program at KTeV}

The KTeV experiment~\cite{ktev} was mainly designed to study the 
Kaon system. The detector was
far (about 94~m) from the production target to ensure mostly $K_{L}$ in the 
neutral beam would reach the detector. However, a copious amount of 
neutrons, and some very high momentum hyperons 
entered the detector along with $K_{L}$'s.
The $\Lambda$ and the $\Xi^0$ were the only two hyperons with lifetimes long 
enough to be observable at the decay volume of the experiment. 

A wide range of topics in hyperon physics is being studied at KTeV. This 
includes but is not limited to:  

\begin{itemize}

\item {Semileptonic (Beta) Decays of $\Xi^0$, $\overline{\Xi^0}$ and $\Lambda$.}
\item {Two Body Radiative Decays of $\Xi^0$: 
\mbox{$\Xi^0 \rightarrow \Sigma^0 \gamma$} and
\mbox{$\Xi^0 \rightarrow \Lambda \gamma$}. }
\item {Three Body Radiative Decays of $\Xi^0$ and $\Lambda$:
\mbox{$\Xi^0 \rightarrow \Lambda \pi^0 \gamma$} and
\mbox{$\Lambda \rightarrow p \pi^- \gamma$}. }
\item {Rare Decays of $\Sigma^0$ (produced from $\Xi^0 \rightarrow \Sigma^0 \gamma$ 
decays): \mbox{$\Sigma^0 \rightarrow \Lambda e^+ e^-$}. } 
\item {Polarization measurements of $\Xi^0$ and $\overline{\Xi^0}$.}
\item {Precision mass measurements of $\Xi^0$ and $\overline{\Xi^0}$.}
\end{itemize}

We made the first observation of some of the above decays, and increased the 
world statistics of the previously observed ones by one to two 
orders of magnitudes.
This discussion will focus on the beta decays of $\Xi^0$ particle. 

\section{The beam and detector}

\begin{figure}[htctb]	
\centerline{\epsfxsize 7.0cm \epsfbox{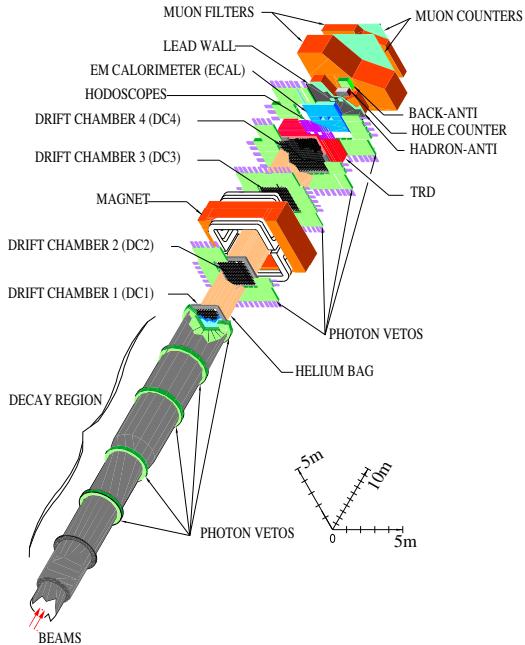}}   
\vskip -.2 cm
\caption[]{The KTeV apparatus in E799 configuration.}
\label{detector}
\end{figure}

An 800~GeV/c proton beam, with 
up to $5\times 10^{12}$~pro\-tons per 19~s Tevatron spill every minute, was
targeted at a vertical angle  of 4.8 mrad  on a 1.1 interaction length
(30~cm) BeO target. A set of sweeping magnets was used to remove the
charged  particles and a set of
collimators defined two nearly parallel neutral beams
that entered the KTeV apparatus (Fig.~\ref{detector})
94~m downstream from the target. The 65~m   vacuum
($\sim$$10^{-6}$~Torr) decay region  extended  to the   first   drift
chamber.

The charged particle spectrometer  consisted of a dipole
magnet  surrounded by four drift
chambers  \mbox{(DC1--4)} with $\sim$100~$\mu$m   position  resolution
in both horizontal and vertical views. To
reduce multiple scattering,  helium filled bags occupied the  spaces between 
the drift  chambers.  In E799-II,  the magnetic field imparted a 
$\pm 205$~MeV/c
horizontal   momentum   component to   charged   particles, yielding a
momentum  resolution    of   $\sigma(P)/P\  =   0.38    \%\,  \oplus\,
0.016\%~P\,$(GeV/c). 

The ($1.9 \times 1.9$~m$^2$)  electromagnetic calorimeter (ECAL)  consisted  
of 3100 pure  CsI crystals. Each crystal was  50~cm  long (27~radiation lengths,
1.4~interaction lengths).  Crystals in the central region ($1.2 \times
1.2$~m$^2$) had a cross-sectional area of $2.5  \times 2.5$~cm$^2$; those in
the outer region, $5 \times 5$~cm$^2$.
After calibration, the ECAL energy  resolution was better than 1\%
for the electron momentum between 2 and 60 GeV.
The position resolution was $\sim$1~mm. We also used the ECAL as the main
particle identification detector. It had a $e/\pi$ rejection of better than 500:1. 

Nine photon veto  assemblies  detected  particles leaving  the  fiducial
volume.   Two scintillator  hodoscopes  in front of the
ECAL were used to  trigger  on  charged particles.  Another scintillator 
plane (hadron-anti), located behind both the  ECAL  and a 10~cm lead
wall, acted as a hadron shower veto. The hodoscopes and the ECAL detectors had
two holes ($15 \times 15$~cm$^2$ at the ECAL) and  the hadron-anti had a single
$64 \times 34$~cm$^2$ hole to let the neutral beams pass through without 
interaction. Charged particles passing through these holes were detected 
by $16 \times 16$~cm$^2$ scintillators (hole counters) located along each 
beam line in the hole region just downstream of the hadron-anti. 


\section{\casb Decay}

\begin{figure}[htctb]	
\centerline{\epsfxsize 4.5in \epsfbox{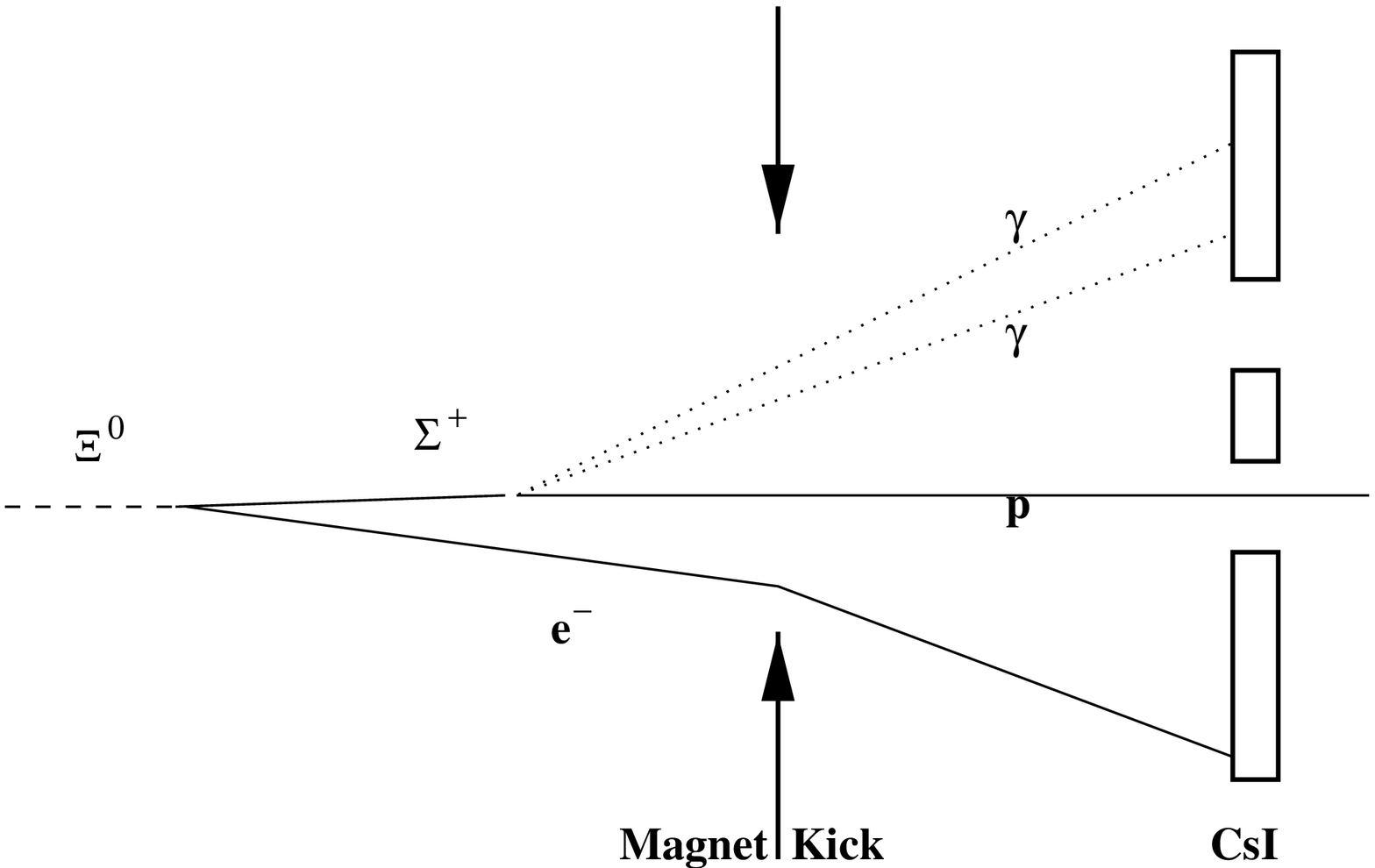}}   
\vskip 0.5 cm
\caption[]{The topology of the decay \casb where 
\sigp and \pizgg}
\label{topology}
\end{figure}

The topology of the decay, \casb~ followed by \sigp (shown in Fig.~\ref{topology}),
is similar to
the dominant $\Xi^0$ decay sequence, \casn followed by
\lamn, which was used for normalization.

Both sequences had a high momentum ($>$100~GeV/c) positive track
(proton) which remained in or near the neutral beam region,
a second lower momentum negative track ($\pi^-$ or  $e^-$),  
and two neutral ({\it i.e.} not associated with any track) ECAL energy clusters 
(photons from a $\pi^0$).  
The beta decay was distinguished  by the presence  of a decay  electron
and by its different vertex structure. 
We used a dedicated {\it beta-trigger} to collect \casb events which was a subset of
a more general {\it hyperon-trigger} with some tighter requirements at the trigger level 
to optimize the number of signal events and reduce the more frequent \lamn events.

To reconstruct a \casb event, we looked at events with two tracks and three in-time
electromagnetic clusters one of which was associated with the negative track.
The  secondary  $\Sigma^+$ decay vertex was located at the point 
along  the stiff proton track  
where the two highest energy neutral ECAL clusters matched the
$\pi^0$ mass. The primary  $\Xi^0$ vertex was then defined at the point
of closest  approach  of  the extrapolated  $\Sigma^+$ path   and  the
negative track. We identified  $e^{-}$'s as negative tracks which
deposited more than 93\%  of their energy in  the ECAL. 
Since the decay product contains a missing neutrino, the reconstructed mass of
$\Sigma^+e^-$ would be broad and below the known mass of $\Xi^0$. Luckily, \casb
was the only event which produced $\Sigma^+$ particles. Therefore, reconstructing
this intermediate particle would be an indirect but confident indication of the
signal. 

In fact, the absence of  a competing two-body decay containing a $\Sigma^+$
($\Xi^0 \rightarrow \Sigma^+ \pi^-$ is not energetically allowed) 
eliminated a major potential background to our signal. The possible 
backgrounds were: (a) \ket, \lamn, \lamb decays with two accidental photons; 
(b) \ketr with one accidental photon; (c) \kef, \kppp; and (d) 
\casn with either
\lamn or \lamb as subsequent decays. Besides trigger requirements and reconstruction
techniques, we applied some quality cuts 
which strongly suppressed these backgrounds. 
Detailed Monte Carlo (MC) studies of the signal and the background events suggested
various cuts based on the topology of the decays, momenta of the decay products,
reconstructed mass and momenta of the parent particles etc.
The primary residual background were \ket, \ketr and \casn followed by 
{{\mbox{$\Lambda \rightarrow p + {\rm anything} $}}}.

\subsection{The BR Measurement}

\begin{figure}[htbp]
  \begin{minipage}[t]{2.5in}
    \leavevmode
    \epsfxsize=2.95in \epsfbox{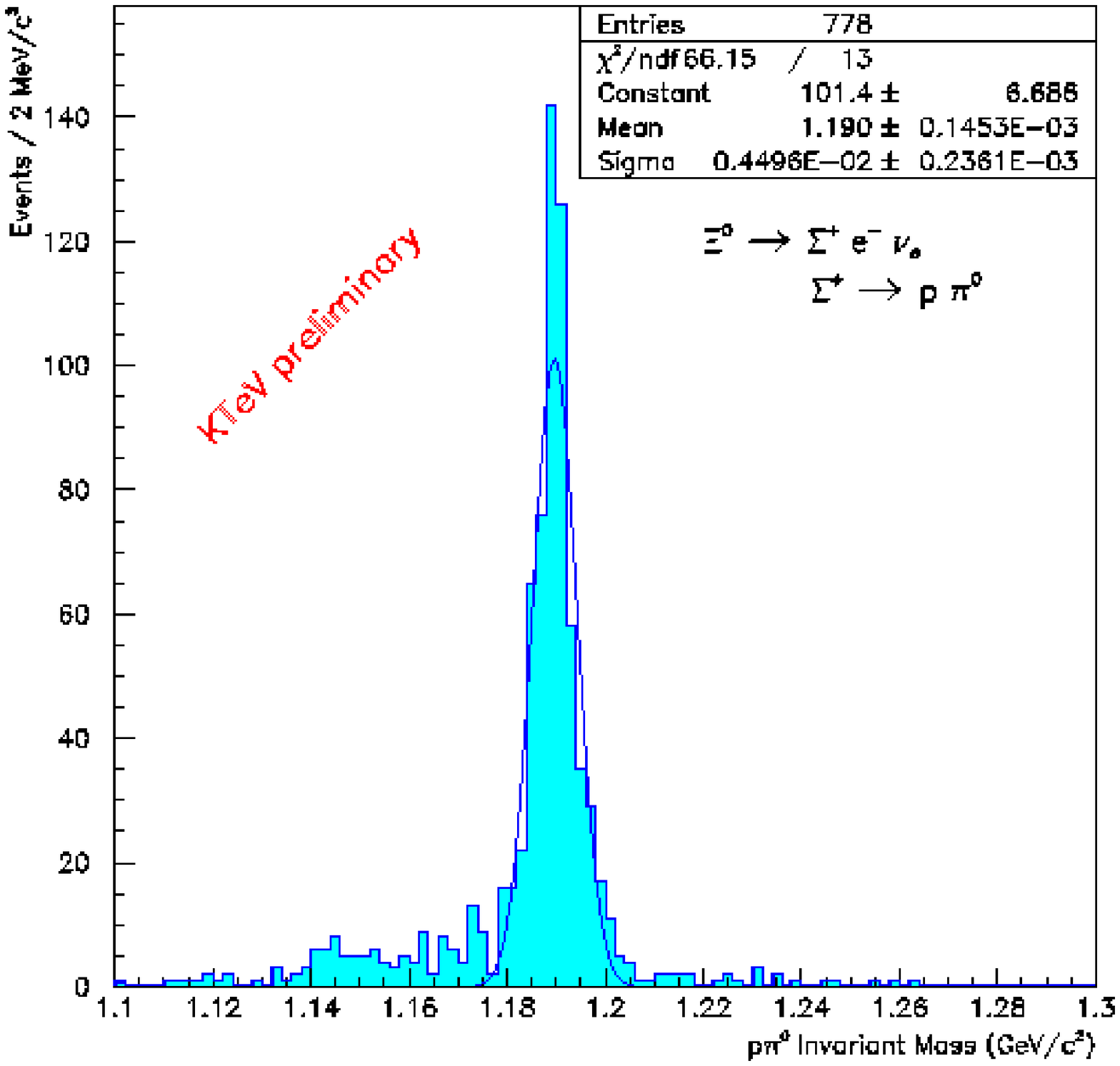}
    \vskip -1.5 cm
    \caption{
    The $p\pi^0$ invariant mass distribution for \casb event candidates from the
    E799-II data taken in Summer of 1997. 
	     }
    \label{signal}
  \end{minipage}
  \hfill
  \begin{minipage}[t]{3.0in}
    \leavevmode
    \epsfxsize=2.95in \epsfbox{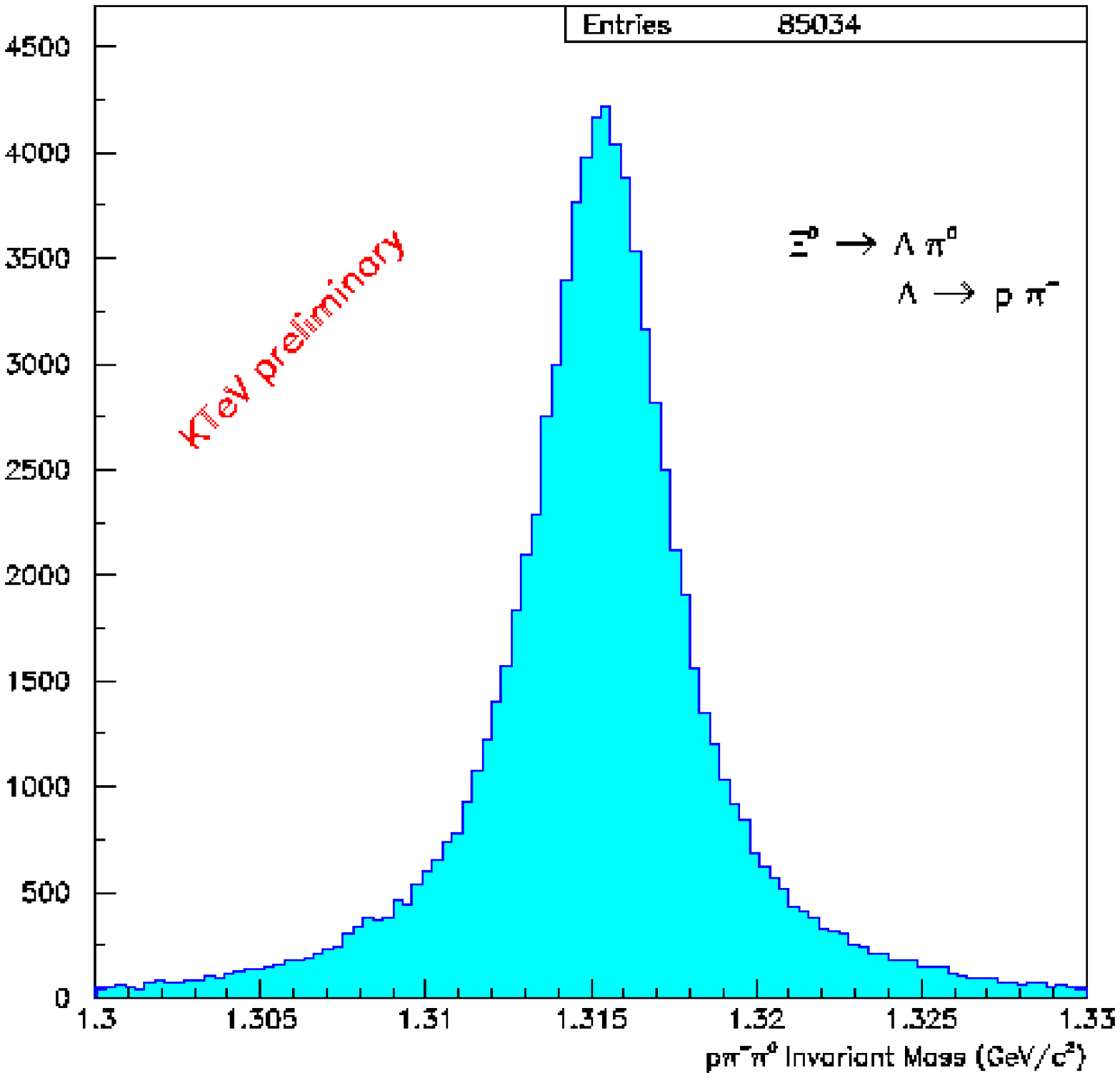}
    \vskip -1.5 cm
    \caption{
     The reconstructed mass of $\Xi^0$ from the decay of \casn used as the flux
     normalization mode for the same data-taking period. 
	     }
    \label{norm}
  \end{minipage}
\end{figure}

Figs.~\ref{signal} and~\ref{norm} show the reconstructed mass of $\Sigma^+$
for the signal mode and $\Xi^0$ for the normalization mode respectively, after
passing all the trigger requirements and analysis cuts. 
The background level in the signal plot is less than 10\% and well 
understood. We collected the signal events from the {\it beta-trigger} which was not 
prescaled, and the normalization events form the nominal {\it hyperon-trigger} which
had a prescale of 50.

The branching ratio of \casb decay normalized to \casn decay can be
calculated from the following relation: 

\begin{center}
$$ 
\frac{BR(\casb\ )}{BR(\casn\ )} = 
\frac{N_{beta}}{N_{norm}} \times \frac{Acc_{norm}}{Acc_{beta}} 
\times 
\frac{BR(\Lambda\rightarrow p\pi^-)}{BR( \Sigma^+\rightarrow p\pi^0)}
\times \frac{Ps(beta-trigger)}{Ps(hyperon-trigger)}
$$
\end{center}

Where 
$N_{beta}$ and $N_{norm}$ are the number of \casb and \casn selected events, 
$Acc_{norm}/Acc_{beta}$=1.54 is the acceptance correction from the Monte Carlo (MC)
simulations of the events, and Ps({\it beta-trigger}) and Ps({\it hyperon-trigger}) are
the prescales of the two triggers.
 
In Fig.~\ref{signal}, there are $626\pm25$ events between 1.175 GeV and
1.205 GeV (within 15 MeV or equivalently $3\sigma$ of the known mass of
$\Sigma^+$ , 1.1894 GeV) on the top of  $60\pm 8$ background events. 
We used a sideband background subtraction method to estimate the level of
background. 
 
Using the values for
 $BR(\Lambda\rightarrow p \pi^-)$ and $BR( \Sigma^+\rightarrow p\pi^0)$  
from~\cite{pdg} we determined the BR:

$$ 
\frac{BR(\casb)}{BR(\casn)} =
(2.54 \pm 0.11_{(stat.)} \pm 0.16_{(syst.)}) \times 10^{-4}
$$

The systematic error has contributions from the trigger inefficiency, 
background 
subtraction and uncertainty in the value of different cuts due to detector 
calibration, and uncertainty in the MC estimation of acceptance. We are planning 
to reduce the systematics in the future. We are also trying to extract a BR
for  $\overline{\Xi^0}$ beta decay based on about 70 candidates~\cite{monniere}.   

\subsection{Form Factor Measurements}

For \casb, the transition amplitude in the V-A theory can be written: 

\begin{eqnarray}  \label{equ:semi_mtx}
{\cal M } =  \frac{G_{F}}{\sqrt{2}}
V_{us} \overline{u}(\Sigma^{+})( V_{\alpha} + A_{\alpha})u(\Xi^{0})
\overline{u_{e}} \gamma^{\alpha} (1+ \gamma_{5})u_{\nu}  
\end{eqnarray}

where $G_{F}$ is the universal weak coupling constant, and $V_{us}$ is the CKM  
matrix element for strangeness changing $\Delta$S=1 decays.
$u(\Xi^{0})$ and $u(\Sigma^{+})$ are the Dirac spinors of the initial and final baryons.
The vector and axial vector currents can be written as 

\begin{eqnarray}
 V_{\alpha} & = & f_{1}(q^2) \gamma_{\alpha} 
+ \frac{f_{2}(q^2)}{M_{\Xi^{0}}}\sigma_{\alpha \beta} 
q^{\beta} +  \frac{f_{3}(q^2)}{M_{\Xi^{0}}}q_{\alpha}  \\
 A_{\alpha} &  = & ( g_{1}(q^2) \gamma_{\alpha} + \frac{g_{2}(q^2)}{M_{\Xi^{0}}}
\sigma_{\alpha \beta} 
q^{\beta} +  \frac{g_{3}(q^2)}{M_{\Xi^{0}}}q_{\alpha} ) \gamma_{5} \
\end{eqnarray}

There are 3 vector from factors $f_1$ (vector), $f_2$ (weak magnetism) and 
$f_3$(induced scalar); plus 3 axial-vector from factors $g_1$ (axial-vector), $g_2$ 
(weak electricity) and $g_3$(induced pseudo-scalar) which are functions of the baryons' 
momentum transfer squared, $q^2$. Time invariance implies that all of them are real.
$f_3$ and $g_3$ are suppressed by the mass of the lepton and can be ignored in
the case of decays to an electron.

The Cabibbo~\cite{cabibbo} theory relates the form factors of different 
Hyperon Semileptonic Decays (HSD)
 to one another by the SU(3)
flavor symmetry assumption. In this limit $g_2$ vanishes (no second-class current) 
and the remaining form factors for e-mode processes at $q^2 = 0$ are written in terms of
only two reduced form factors F and D which are the free parameters in this model.
For Cascade beta decay $f_1(0)=1$ and $g_1= F + D$, similar to the well studied 
neutron beta decay. Thus, in the flavor symmetric quark model, differences 
between these two decays arise only from the differing particle masses and their  
CKM matrix elements.

We are measuring the $g_1/f_1$ ratio at KTeV by looking at the 
electron-proton asymmetry in the rest frame 
of $\Sigma^+$. With an order of 1000 events, $ g_{1} / f_{1} $ can be measured to 
about 0.2. This ratio $g_1/f_1 = 1.2670 \pm 0.0035$ for the neutron beta decay.
This analysis is in progress and the results will be announced soon.

The total decay rate is also a function of $f_1$ and $g_1$, which can be calculated 
from the BR of the decay and the measured lifetime of $\Xi^0$. Hence, these two 
measurements at KTeV can provide a good test of the
SU(3) symmetry assumption and either verify or rule out several theoretical models which
predict the values of these form factors based on SU(3) symmetry breaking assumptions.  

\section{BR measurement of \casmub Decay}

This decay is the muonic channel of $\Xi^0$ beta decay. 
For this decay the contributions from the f$_{3}$ and g$_{3}$ form factors may no 
longer be considered negligible~\cite{linke}. 
Because of a smaller available phase space, the decay rate of this mode is about two 
orders of magnitude smaller than that of \casb decay and therefore more challenging 
to observe. 

For \casmub, We performed a similar analysis as in \casb.
Similar or very close selection criteria were applied,
except the requirement of identifying a muon instead of an electron. 
Various analysis cuts were adjusted to account for the greater mass of muons compared to
electrons.

\begin{figure}[htctb]	
\centerline{\epsfxsize 3.0in \epsfbox{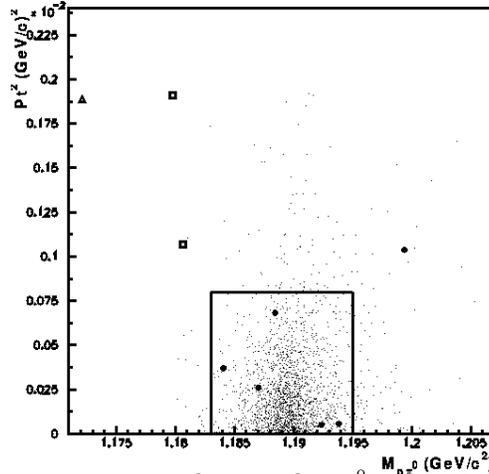}}   
\vskip -1.5 cm
\caption{$\Xi^0$ reconstructed transverse momentum squared 
versus the $p\pi^0$ invariant mass for \casmub event candidates.
The plain circles are data. Dots are MC simulation of the 
decay. Triangles are Kaon background (from opposite sign data),
and open squares are MC
simulated $\Xi^0 \rightarrow \pi^0 \Lambda \rightarrow p 
\pi^- \rightarrow \mu^- \overline{\nu}_\mu$ background events.
Superimposed is a box that contains 90\% of the simulated signal events.}
\label{casmuon}
\end{figure}

To select muons, hits in the muon counters were required in combination with
almost no energy deposited by the muon in the calorimeter and no hadronic showering
in the back of the calorimeter. In addition to usual kinematic criteria, the 
$\pi^-\mu^+\pi^0$ reconstructed mass was required to be greater than $0.49 GeV/c^2$ 
to remove 
$K^0 \rightarrow \pi^0 \pi^+ \pi^-$ with $\pi^-\rightarrow\mu^-\overline{\nu}_\mu$
background. The remaining effects of this background have been studied with
wrong sign events since anti-hyperons were suppressed by a factor of 10-12 at production. 
Mass cuts were also used to remove most of the background of \casn events with
\lamn when the $\pi^-$ either decays in flight or fakes a muon. Finally in all the 
data taken, five events remain with an estimated background of 0 events 
in the 90\% confidence level box as can be seen in Fig.~\ref{casmuon}. 
The single event outside of the box is also
consistent with expectations for the background.
The first observation of this decay mode based on these events was presented
earlier~\cite{monniere}. 
For the BR calculations we used \casb decay~\cite{paper} as the flux
normalization mode.
Based on the five observed events, the BR is measured:

$$ 
BR( \Xi^0 \rightarrow \Sigma^+ \mu^- \overline{\nu_e} ) 
= (2.6^{+2.7(stat.)}_{-1.7(stat.)} \pm 0.6_{(syst.)}) \times 10^{-6}
$$

in good agreement with the SU(3) symmetry prediction of 2.20~$\times$~10$^{-6}$.
The main source of systematic error comes from the uncertainty of the BR of the 
normalizing mode, and uncertainty in the background estimation. 

\section{Conclusion}

We presented the latest results on the measurement of the BR for the two beta
decays of $\Xi^0$ from the KTeV data. They are both in  agreement with the 
Cabibbo model based on SU(3) flavor symmetry assumption for HSD, within the errors. The first 
form factor measurement of \casb will be finalized soon. 
KTeV is approved to run in 1999 and we expect to triple our statistics during the
upcoming data-taking period.

\end{document}